# Direct laser printing of thin-film polyaniline devices

M. Kandyla[1,a], C. Pandis[1,b], S. Chatzandroulis[2,c], P. Pissis[1,d], I. Zergioti[1,e]

[1]*Physics Department, National Technical University of Athens, 9 Iroon Polytechniou Street, Zografou 15780, Athens, Greece*

[2]*Institute of Microelectronics, NCSR Demokritos, Agia Paraksevi 15310, Athens, Greece*

**Abstract:** We report the fabrication of electrically functional polyaniline thin-film microdevices. Polyaniline films were printed in the solid phase by Laser Induced Forward Transfer directly between Au electrodes on a Si/SiO$_2$ substrate. To apply solid-phase deposition, aniline was *in situ* polymerized on quartz substrates. Laser deposition preserves the morphology of the films and delivers sharp features with controllable dimensions. The electrical characteristics of printed polyaniline present ohmic behavior, allowing for electroactive applications. Results on gas sensing of ammonia are presented.

[a] Email: kandyla@central.ntua.gr, tel: +30 2107273826, fax: +30 2107723025
[b] Email: pandis@central.ntua.gr, tel: +30 2107721712, fax: +30 2107722932
[c] Email: stavros@imel.demokritos.gr, tel: +30 2106503271, fax: +30 2106511723
[d] Email: ppissis@central.ntua.gr, tel: +30 2107722986, fax: +30 2107722932
[e] Email: zergioti@central.ntua.gr, tel: +30 2107723345, fax: +30 2107723338

# 1. Introduction

Conducting polymers offer an attractive alternative to their inorganic counterparts due to properties such as flexibility, low cost, light weight, and ease of processing. Polyaniline is one of the most commonly used conducting polymers due to its high conductivity, unique redox properties, excellent environmental stability, and ease of preparation. Applications based on polyaniline include chemical and biological sensors [1,2], batteries [3], supercapacitors [4], transistors [5], and conducting coatings [6].

The practical use of polyaniline is hindered by its mechanical properties and its poor solubility in most common organic solvents. In order to overcome this difficulty, various methods have been developed, including blending polyaniline with another polymer [7,8], incorporating side groups into the main chain [9], employing a soluble graft copolymer [10], forming polyaniline dispersions [11,12], and electrochemically polymerizing aniline in a polymer matrix [13]. Therefore, the development of a simple and reliable method for patterning devices from conventional conducting polyaniline is still a challenge.

We present results on the patterned deposition of thin polyaniline films, prepared with a method suitable for various applications, directly from the solid phase, without the use of solvents, electrochemistry, or additional polymers, by using Laser Induced Forward Transfer [14,15]. We are able to control the dimensions of the deposited films and achieve spatial resolution of a few μm, thus enabling the practical use of conducting polyaniline, synthesized with a 'standard' procedure [16,17], in microdevices. The deposited films maintain the electrical properties of conducting polyaniline, which is important because many applications of polyaniline are based on the electroactive switching properties of this polymer. Laser Induced Forward Transfer is a dry, contactless, and versatile technique, which enables the controlled transfer of a thin film from a transparent carrier to a receiving substrate. Realizing the transfer while the material carrier and the receiving substrate move with respect to each other, allows for the fabrication of 2D and 3D patterns. In this work, a nanosecond laser is used to transfer selected areas of controllable dimensions from polyaniline films, grown *in situ* on transparent quartz carriers, onto Si/SiO$_2$ substrates bearing Au electrode pairs. This way, electrically functional devices are fabricated based on polyaniline. As an example, we present preliminary results on the gas sensing capability of the deposited films, taking advantage of the well-known sensing properties of polyaniline [18,19] for ammonia detection.

## 2. Experimental

Anilinium chloride and ammonium peroxodisulphate were purchased from Panreac, Spain and used as received. For the synthesis of polyaniline films we followed the *in situ* method described by Stejskal *et al*. [20]. 0.2 M anilinium chloride was oxidized with 0.2 M ammonium peroxodisulphate in 1.0 M HCl at 5 ºC. Quartz substrates were covered by adhesive tape on one side and placed in the reaction vessel, where they stayed overnight. After the polymerization was over, the substrates were well rinsed with dilute HCl to remove the adhering polyaniline precipitate, and dried. The quartz samples covered by a thin polyaniline film were then separated from the adhesive tape and used as carriers for Laser Induced Forward Transfer.

Electrode-bearing Si/SiO$_2$ substrates were prepared using an n-type (resistivity 1-10 Ωcm), 100-mm diameter Si wafer. The wafer was first oxidized using dry oxidation to form a 200-nm thick oxide on its surface. Subsequently, a thin Au film (60 nm) was deposited and patterned using lift-off techniques, in order to form the electrode pairs. Finally, the wafer was diced into small samples.

For morphological characterization, an Atomic Force Microscope (AFM) (diInnova, Veeco) and a Field Emission Scanning Electron Microscope (FE-SEM) (230 Nova NanoSEM, FEI Company) were used.

In order to test the prepared structures as gas sensors for ammonia detection, the devices were wire bonded on a printed circuit board (PCB) and placed in an air-tight test chamber with controllable atmosphere, allowing the measurement of electrical resistance. The resistance of the devices was monitored in room temperature using a 2400 Source-Meter (Keithley). The sensor response was determined as a relative change of the resistance R of the sensor exposed to the analyte (ammonia gas) compared to the initial value $R_o$ when the sensor was exposed to a dry nitrogen flow: $((R - R_o)/R_o) \times 100\%$. The desired analyte concentration in the test chamber was established by controlling the mixing of nitrogen and ammonia by means of two mass flow controllers (Bronkhorst BV).

## 3. Results and Discussion

We performed Laser Induced Forward Transfer of polyaniline on electrode-bearing Si/SiO$_2$ substrates for device fabrication and on glass substrates for characterization. In order to carry out Laser Induced Forward Transfer, the receiving substrate is placed underneath the transparent film carrier, facing the thin film at a very short distance. The laser beam is then focused through the transparent carrier at the thin-film/carrier interface and transfers the irradiated area of the film onto the substrate. The experimental apparatus is described in detail elsewhere [21,22]. As a laser source we use the 4$^{th}$ harmonic of a Q-switched Nd:YAG laser system (4 ns pulse duration, 266 nm wavelength). The laser beam is propagated through a variable rectangular mask, which selects the middle part of the beam and reduces its size. The selected part of the beam is then imaged onto the polyaniline film surface, through a system of lenses and a 15× microscope objective. The substrate is placed parallel to the polyaniline carrier, at a distance less than 50 μm. By focusing one laser pulse on the polyaniline surface, we are able to deposit a square polyaniline spot on the substrate, for device fabrication or characterization. Figure 1 shows the absorbance of the polyaniline films we prepare for laser printing. The spectrum is characteristic for polyaniline in the conducting state [23]. At 266 nm, which is the wavelength of the laser system, the absorbance of the films is high enough for direct laser transfer without the need for intermediate sacrificial absorbing layers [24,25]. At the same time, the absorbance is not too high to damage the material's electric properties of interest, as we will see below.

Figure 2a shows an optical microscope image of polyaniline spots deposited by Laser Induced Forward Transfer between Au electrodes lying on a Si/SiO$_2$ substrate. The channel width between the electrodes is 40 μm and the spots dimensions are 55 × 55 μm$^2$. The spots were transferred directly from the solid phase from a bigger quartz/polyaniline carrier. Profilometer measurements (not shown here) reveal the thickness of the *in situ* polymerized film on quartz is 200 nm, which is the same as the thickness of the spots after laser deposition [26]. The measured thickness of 200 nm is typical for polyaniline films grown with the method and the conditions employed in this work [20]. The deposition of each polyaniline film was achieved by one laser pulse for an incident fluence of 160 mJ/cm$^2$. A scanning electron microscope (SEM) image of a laser-deposited polyaniline spot is shown in Fig. 2b, showing the area between the Au electrodes, which is entirely covered by conducting polyaniline. This ensures there is a conducting path between the two electrodes, which can be used as a microdevice for electroactive applications of polyaniline.

In order to investigate the effect of laser printing on the morphology of the films, we employed atomic force microscopy (AFM). Figure 3a shows an AFM image of a polyaniline film, polymerized *in situ* on a quartz carrier before laser printing. The globular morphology of the film is typical for polyaniline synthesized by the method employed in this work [27]. Figure 3b shows an AFM image of laser-printed polyaniline from the same quartz carrier. The rms roughness of the polyaniline film is 0.12 μm before laser printing and 0.10 μm after laser printing, therefore Laser Induced Forward Transfer does not alter the surface roughness of the deposited films significantly. A certain degree of surface roughness is useful for sensing applications because it increases the active area of the sensor. Polyaniline devices like the ones presented in this work may be used successfully as electrochemical microsensors. Figure 3c shows an AFM image of the edge of a laser-deposited polyaniline film. We can see the laser cut is precise and the edge of the printed spot is sharply defined.

We measured the conductivity of polyaniline before and after laser deposition. The conductivity of polyaniline *in situ* polymerized on the quartz carrier was found 0.56 S/cm$^2$, while after laser deposition the conductivity was found 0.04 S/cm$^2$, showing a decrease by one order of magnitude. It is known that deprotonation of polyaniline from emeraldine salt to emeraldine base, which results in conductivity decrease, occurs at modest temperatures [28,29]. Additionally, polyaniline is known to undergo backbone degradation, crystallinity decrease, and morphological changes such as cross-linking and chlorination of benzene rings at elevated temperatures, which also result in conductivity decrease [30]. A separate study aiming at understanding the origin of conductivity decrease during laser printing of polyaniline is under way. For electroactive polyaniline applications, the conductivity decrease we observe is acceptable in most cases. Figure 4a shows average I-V characteristics of polyaniline samples, laser-printed between Au electrodes on Si/SiO$_2$ substrates. The measurements were taken with a HP 4140B pico-amperometer equipped with microcontacts at room temperature (~20 ºC). All samples are conducting and present ohmic behavior, which is essential for their use in electroactive applications. The electric response of the samples remains linear over orders of magnitude of current and voltage, as we can see in Fig. 4b.

Figure 5 shows the response ((R - Ro)/Ro) × 100% of the prepared device for consecutives cycles of exposure to ammonia flow at various concentrations, from 500 – 2000 ppm. The typical increase of polyaniline resistance when exposed to ammonia was observed, as well as the recovery of the response after the removal of the analyte. This effect is well known in the literature [19,31] and is attributed to the deprotonation of PANI due to the interaction with ammonia, leading to an increase of electrical resistance [32]. The inset shows the maximum signal as a function of ammonia concentration (also known as the calibration curve), which indicates the linearity of the sensor response in the measured concentration range. Additional measurements (not shown here) indicate good reproducibility of the response of the prepared devices to ammonia. The sensing quality shown here allows for the practical use of similar devices for many gas sensor applications [33].

## 4. Conclusion

The laser deposition method presented in this work overcomes many of the disadvantages of polyaniline. Its poor mechanical properties and lack of solubility in organic solvents do not affect the Laser Induced Forward Transfer, which depends mainly on the optical properties of the material and allows for solid-state processing. Nozzle clogging, which is an inherent obstacle for inkjet printing of polyaniline, is not a concern in the experiments presented here. Additionally, even though the *in situ* polymerization used in this work takes place in the presence of excess HCl, the polymer film is subsequently removed from the solution, therefore the increased pH of the polymerizing environment is not a concern for applications (*e.g.* biological applications). Laser deposition does not damage the electrical properties of polyaniline films and offers a reliable method for the fabrication of electrically functional microdevices. An array of such devices can be easily fabricated on the same substrate, for multipurpose applications. Laser printing is critically affected by the thickness of the donor material on the transparent carrier. The thickness of a solid donor does not suffer from evaporation and surface tension effects that tend to deteriorate the uniformity and stability of the thickness of liquid donors. The solid donors prepared in this work provide reproducible laser deposition results. Furthermore, by varying the polymerization temperature, we can control the thickness of the polyaniline donors [20] and eventually the thickness of the deposited films. Finally, printing from the solid phase does not involve the deposition of solvents on the substrate, making this method highly suitable for sequential deposition of multilayer structures and all-organic devices.

The sensing results presented here, offer an example of the applicability of Laser Induced Forward Transfer for the preparation of polyaniline microdevices. As this work was focused on the patterning process of the microdevices using a simple conventional method for the synthesis of polyaniline, future work will concentrate on the optimization of the synthetic route (conducting polymer or nanocomposite), in order to enhance the sensing properties of the final device.

## Acknowledgments

The research leading to these results was supported by the European Commission under a Marie Curie International Reintegration Grant, Seventh Framework Program, grant agreement 224790.

**Figure Captions**

**Figure 1**: Absorbance of *in situ* polymerized polyaniline films. The arrow indicates the wavelength of the laser system.

**Figure 2**: (a) Optical microscope image of polyaniline spots deposited by Laser Induced Forward Transfer between Au electrodes on a Si/SiO$_2$ substrate. (b) Scanning electron microscope image of a laser-deposited polyaniline spot, showing the area between the electrodes.

**Figure 3**: Atomic force microscope image of a polyaniline film (a) before laser deposition and (b) after laser deposition. (c) The edge of a laser-deposited polyaniline film.

**Figure 4**: Current-voltage (I-V) characteristics of polyaniline films laser-deposited between Au electrodes on a Si/SiO$_2$ substrate, on (a) a linear and (b) a logarithmic scale.

**Figure 5**: Response of the polyaniline device in the presence of ammonia gas for various concentrations. The inset shows the maximum response as a function of ammonia concentration.

# Figures

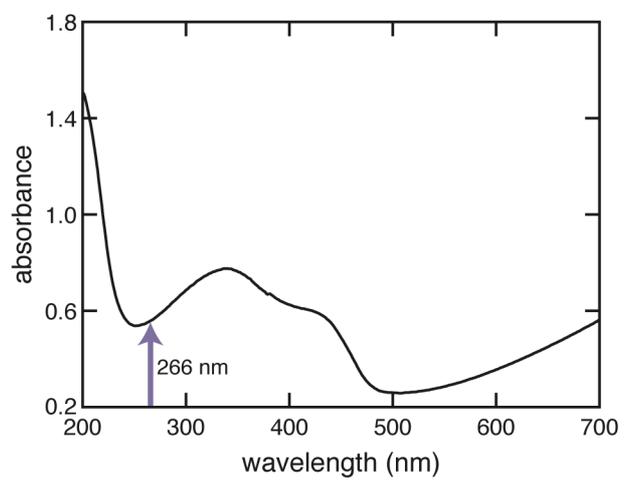

Figure 1

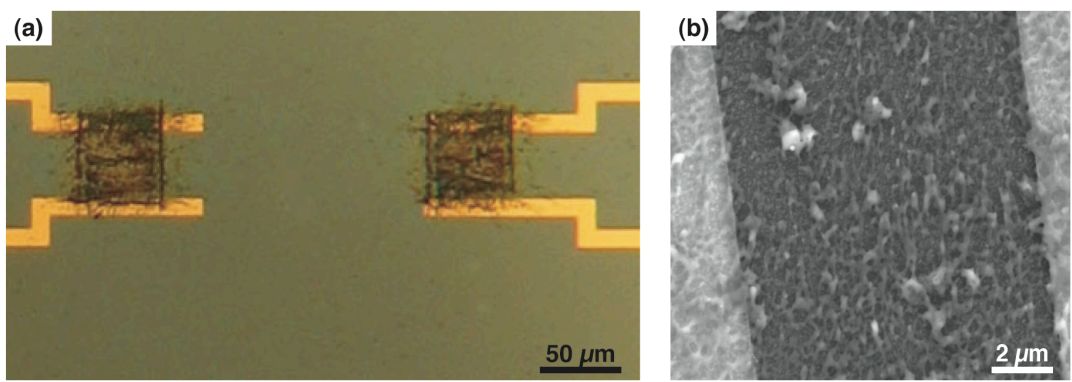

Figure 2

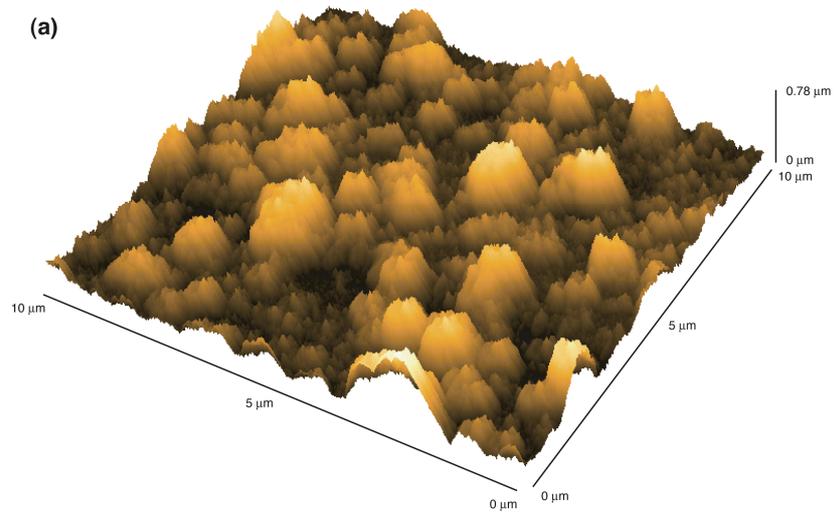

rms roughness: 0.12 µm

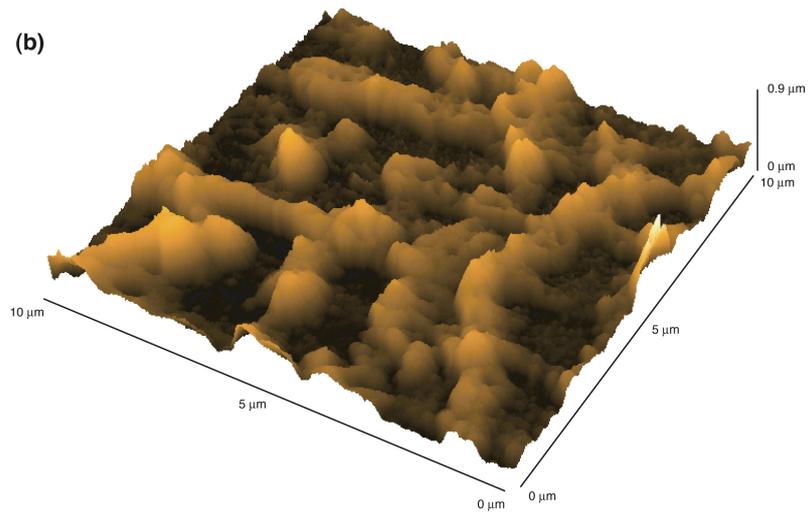

rms roughness: 0.10 µm

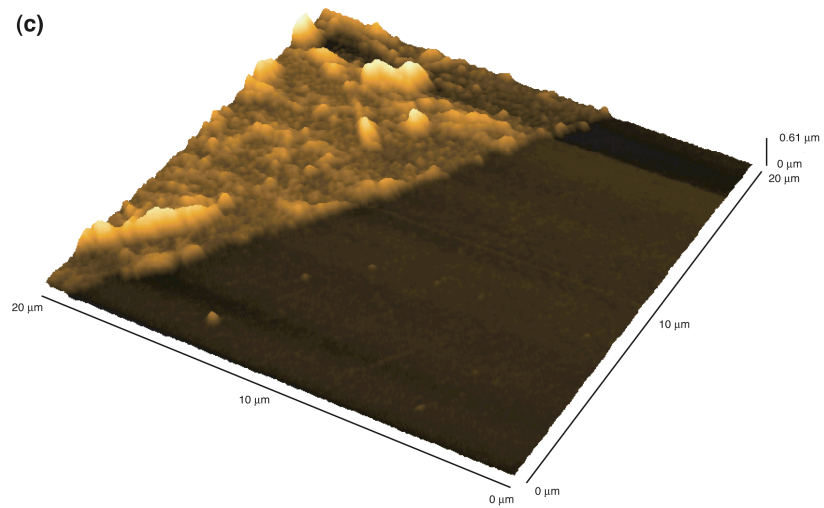

— 15 —

**Figure 3**

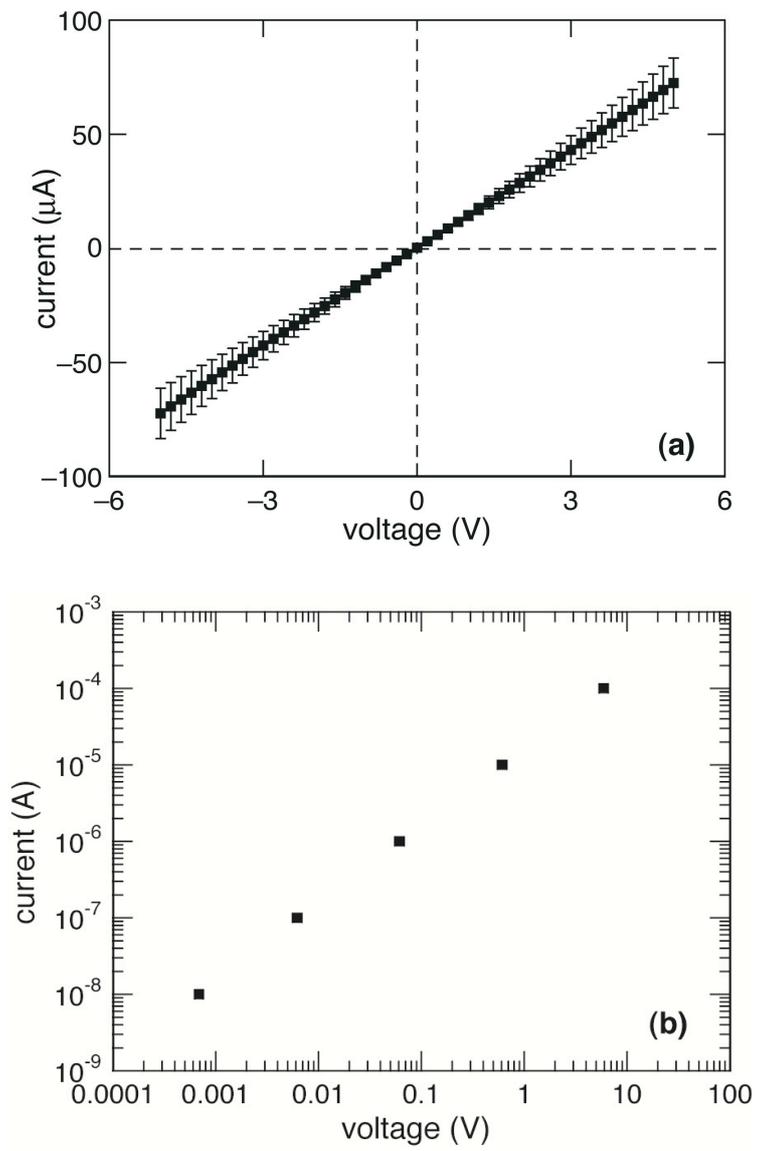

**Figure 4**

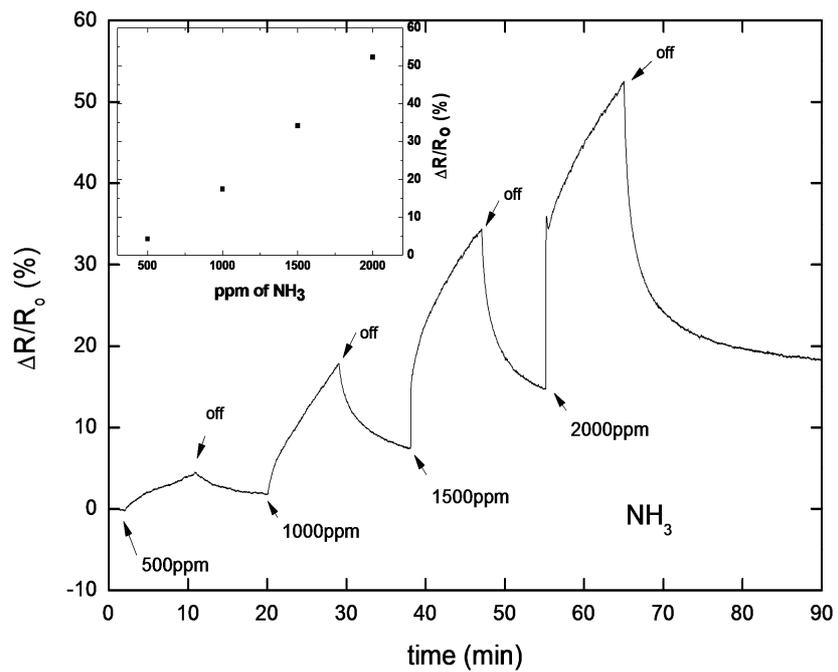

**Figure 5**